\documentclass[]{aa}                                                   
\usepackage{psfig}                                                              
\begin{document}                                                                
                                                                           
\title{Distance of the Hypernova SN 2002ap via the Expanding
Photosphere Method  \footnote[1]{                     
Based on observations obtained at the David Dunlap Observatory,
University of Toronto}}                                                      
                                                                                
\author{J. Vink\'o \inst{1,6} \and   
R. M. Blake \inst{2} \and
K. S\'arneczky \inst{1,6} \and    
B. Cs\'ak \inst{1,6} \and   
G. F\H ur\'esz \inst{3,6} \and   
Sz. Csizmadia \inst{3} \and       
L.L. Kiss \inst{4,5}
Gy.M. Szab\'o \inst{4,6} \and
R. Szab\'o \inst{3} \and 
H. DeBond \inst{2} \and
M.M de Robertis \inst{2} \and
J.R. Thomson \inst{2} \and
S.W. Mochnacki \inst{2}                                                    
}

\institute{Department of Optics \& Quantum Electronics, University of Szeged,   
   POB 406, Szeged, H-6701 Hungary \and                                            
David Dunlap Observatory, University of Toronto, POB 360,
   Richmond Hill ON, Canada L4C 4Y6 \and                            
Konkoly Observatory of the Hungarian Academy of Sciences, POB 67,               
   Budapest, H-1525 Hungary \and                                                   
Department of Experimental Physics, University of Szeged, D\'om t\'er 9., Szeged
   H-6720 Hungary \and
School of Physics, University of Sydney, Australia \and    
Guest Observer, Piszk\'estet\H o Station, Konkoly Observatory, Hungary
} 
                                                                                
\titlerunning{The distance of SN 2002ap}                                             
\authorrunning{J. Vink\'o {\it et al.}}                                         
\offprints{vinko@physx.u-szeged.hu}                                             
\date{}                                                                         
                          
\abstract{New optical photometric and spectroscopic data of the bright
hypernova SN 2002ap are presented. The obtained $BVRI$ light curves
as well as the optical spectra agree well with other published data series. 
The distance has been inferred by applying
the Expanding Photosphere Method for the data around maximum.
The derived $6.7$ Mpc is in good agreement with recent photometric
distances of M74. However, the total (random plus systematic) uncertainty
of the EPM-distance is at least $\pm 4.5$ Mpc (about 70 \%).  
The physical parameters of the SN 
have been determined via simplified analytic models of the light
and velocity curves.
It is confirmed that SN 2002ap was a less energetic 
hypernova, the kinetic energy was 4 - $8 \cdot 10^{51}$ erg, and the
reddening-free absolute bolometric magnitude reached 
$-16.63$ mag (corresponding to $L_{bol}~ = ~3.47 \cdot 10^8 L_{\odot}$),
about 2 magnitude less than the prototype hypernova SN~1998bw.
}

\maketitle


\section{Introduction}

The idea that core-collapse supernovae (SNe) may produce 
observable gamma-ray bursts, has received many theoretical
as well as observational attention since the discovery of
SN~1998bw within the error box of GRB~980425 (\cite{b1}).
Recently, the optical afterglow of GRB~030329 that showed
SN~1998bw-like spectrum (SN~2003dh, \cite{b2}; \cite{b3}; \cite{b4};
\cite{lipkin})
provided a very convincing evidence of the supernova - GRB connection. 
With GRB~031203 - SN~2003lw (\cite{galy2}) there are now three low-redshift
($z < 0.3$) GRBs that are all associated with SNe. 
The spectra of these SNe were characterized by shallow, very broad lines and
a very blue continuum at the early phases. Usually, their spectroscopic
classification is Ic, although they do not really resemble other
Type Ic SNe. The very broad lines in the early spectra are
indicative of high expansion velocity, up to 40,000 kms$^{-1}$.
Such SNe are now referred as ``hypernovae'', because of their
high expansion velocity as well as large kinetic energy released
by the explosion (\cite{b5}). Actually, some model computations 
(\cite{hoflich}) showed that such high expansion velocities 
can also be consequences of an asymmetric explosion seen 
at different viewing angles rather than extremely high 
intrinsic kinetic energy. 

SN~2002ap, discovered by Y. Hirose on Jan. 29, 2002 
(\cite{b6}), turned out to be a very interesting event in
many respects. First, it occured in M74 (NGC 628), a nearby spiral galaxy
($d \approx 7$ Mpc), thus, it is one of the closest SNe
observed ever. Second, it was discovered well before maximum,
and the early phase spectra were very similar to those of
SN~1998bw. Therefore, SN~2002ap was quickly classified as
a hypernova (\cite{b7}; \cite{b8}; \cite{b9}). Third, despite
the expectations, SN~2002ap could not be associated with
any detected GRBs (\cite{b12}). Some authors (\cite{leo}) 
suggested that SN 2002ap should be classified as 
``Peculiar Type Ic'' SN instead of hypernova, because 
in this case the latter term can be reserved to those SNe that 
actually produce GRBs.

In spite of the lack of associated GRB,
SN~2002ap has been extensively followed-up in many bands
extending from radio (\cite{b10}) to X-ray (\cite{b11}; \cite{soria}). 
Optical spectroscopy and/or photometry were reported 
in many papers  
(\cite{galy}; \cite{bori}; \cite{cook}; \cite{pande};
\cite{kinu}; \cite{yosh}; \cite{foley}). 
Based on optical spectrophotometry, significant continuum
polarization was detected (\cite{kawa}; \cite{leo}; 
\cite{wang}), suggesting an asymmetric explosion.

From these observations, a more-or-less consistent physical
picture was revealed, containing a relatively low-mass 
progenitor ($M = 2 - 5 ~M_{\odot}$), exploding with 
$E \approx 4 - 10 ~ 10^{51}$ ergs kinetic energy,
synthesizing 0.07 - 0.1 $M_{\odot}$ of $^{56}$Ni 
(\cite{mazz}; \cite{maeda}; \cite{foley}). The original
mass of the progenitor was estimated as 
$M_{ZAMS} \approx 20 - 25 ~ M_{\odot}$. However, the lack of
the detection of the progenitor on previous optical and near-IR
frames of M74 (\cite{smartt}) may imply a lower mass and/or 
lower luminosity. 

All of these physical parameters are based on the assumed distance
to the host galaxy of SN~2002ap. There are two modern distance
estimates of M74 in the literature. \cite{shari} derived
$\mu_0 = 29.32$ mag as a reddening-free distance modulus, based on
the luminosity of the brightest blue supergiants. The Sharina et al.
result confirms the distance determined by  \cite{sohn} ($\mu_0 = 29.3$), 
although  Sohn \& Davidge claimed that blue supergiants are unreliable
distance indicators and they preferred the use of the brightest 
red supergiants for this purpose.  However, both groups assumed
the total absorption in $B$ as $A_B = 0.13$ mag based on the map
of \cite{buhe}. The more recent reddening map of \cite{sfd} predicts
$A_B = 0.301$ mag in the direction of M74, which reduces the above
distance moduli to $\mu_0 = 29.15$ mag, corresponding to $d = 6.8$ Mpc.

Surprisingly, the various attempts to determine the distance of M74
have led to very inconsistent results. For example, \cite{st} derived
$\mu_0 = 31.46$ mag from the sizes of HII-regions, while \cite{bott}
inferred $\mu_0 = 26.60$ from Tully-Fisher relation. Thus, there is 
an order of magnitude disagreement between the distances determined by
different methods.  
 
In all papers cited above, the distance of SN~2002ap was assumed 
to be 7.3 Mpc, corresponding to the original distance modulus of
\cite{shari}. This value itself may introduce a 0.2 mag systematic
uncertainty in the derived absolute magnitudes of the SN (which then
incorporates into the other inferred physical parameters), because
this distance is based on the earlier reddening map. Moreover, the
large scatter in the early M74 distances may further complicate the
situation. A new, independent distance measurement of M74 based on
SN 2002ap itself seems to be useful to clarify the picture.
 
In this paper we present a new distance estimate of SN 2002ap and
M74 via the Expanding Photosphere Method (EPM). For this purpose
we obtained new optical spectrocopic and photometric measurements
of SN~2002ap, from $t=-4$ days up to $t=+291$ days. These are 
described in Section 2. The distance estimate is discussed in
Section 3. Section 4 contains the analysis
of the bolometric light and radial velocity curves, based on
a simple analytic model. Section 5 summarizes our results.

\section{Observations and data reduction}

\subsection{Photometry}
The photometric observations were started on Jan 31, 2002, just after
the discovery of SN~2002ap. Most of the data were obtained with
the 60/90 cm Schmidt-telescope of Konkoly Observatory (installed
at Piszkesteto Mountain, Hungary) equipped with a 1536x1024 
Photometrics camera. Some CCD frames were also collected
using the 1 m RCC telescope at Konkoly and the 40 cm Cassegrain
at Szeged Observatory (see \cite{vinko2} for the technical 
details of these instruments).

The CCD frames were reduced using the standard way in 
{\it IRAF}\footnote{{\it IRAF} is distributed by 
NOAO which is operated by the Association of Universities for
Research in Astronomy (AURA) Inc. under cooperative agreement with the National 
Science Foundation}. Comparison stars were selected in the
vicinity of the SN (Fig.1), and differential magnitudes
were determined via aperture photometry. Because the SN occured
outside the visible region of its host, the background was
not affected significantly by the light of the host galaxy. 
Thus, aperture photometry could be easily applied.   

\begin{figure}
\begin{center}
\psfig{file=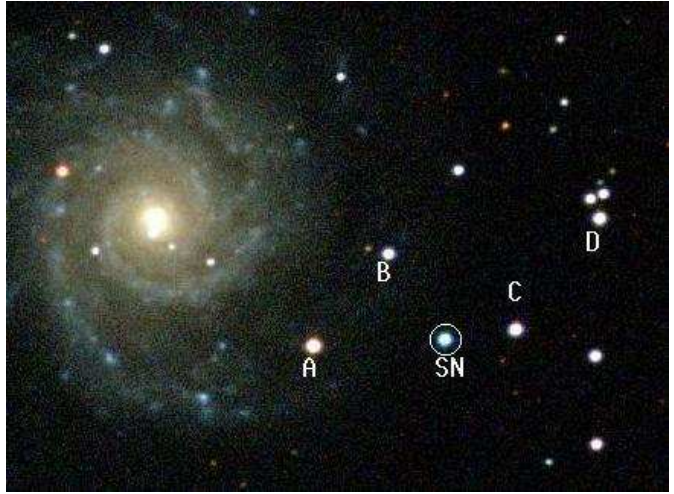,width=8.8cm}
\caption{SN~2002ap and the local comparison stars in the field of M74.}
\end{center}
\end{figure}

The instrumental magnitudes were transformed into the standard
system using the equations given by \cite{vinko1}. The comparison
stars were calibrated via standard stars in the field of M67.
Then, the standard magnitudes were compared with those given by
Henden (\cite{galy}) and Foley et al. (2003). 
The three datasets showed acceptable agreement within 0.03 mag (Fig.2),
except for the brightest star (A), where the differences were slightly higher.
The final adopted magnitudes of our comparison stars are determined 
as the average of the three standard sequences. 
These are collected in Table~1. The uncertainties are estimated
from the standard deviation of the averaged data. 

\begin{table}
\caption{Final adopted magnitudes of local standard stars. Errors
are given in parentheses.}
\begin{tabular}{lcccc}
\hline
\hline
Star & $V$ & $B-V$ & $V-R$ & $V-I$ \\
\hline
A & 12.91(0.05) & 1.24(0.06) & 0.78(0.02) & 1.46(0.04) \\
B & 13.80(0.02) & 0.72(0.01) & 0.42(0.02) & 0.81(0.03) \\
C & 13.08(0.03) & 0.76(0.02) & 0.45(0.01) & 0.89(0.03) \\
D & 13.24(0.02) & 0.77(0.02) & 0.44(0.01) & 0.87(0.02) \\
\hline
\end{tabular}
\end{table}
  
\begin{figure}
\begin{center}
\leavevmode
\psfig{file=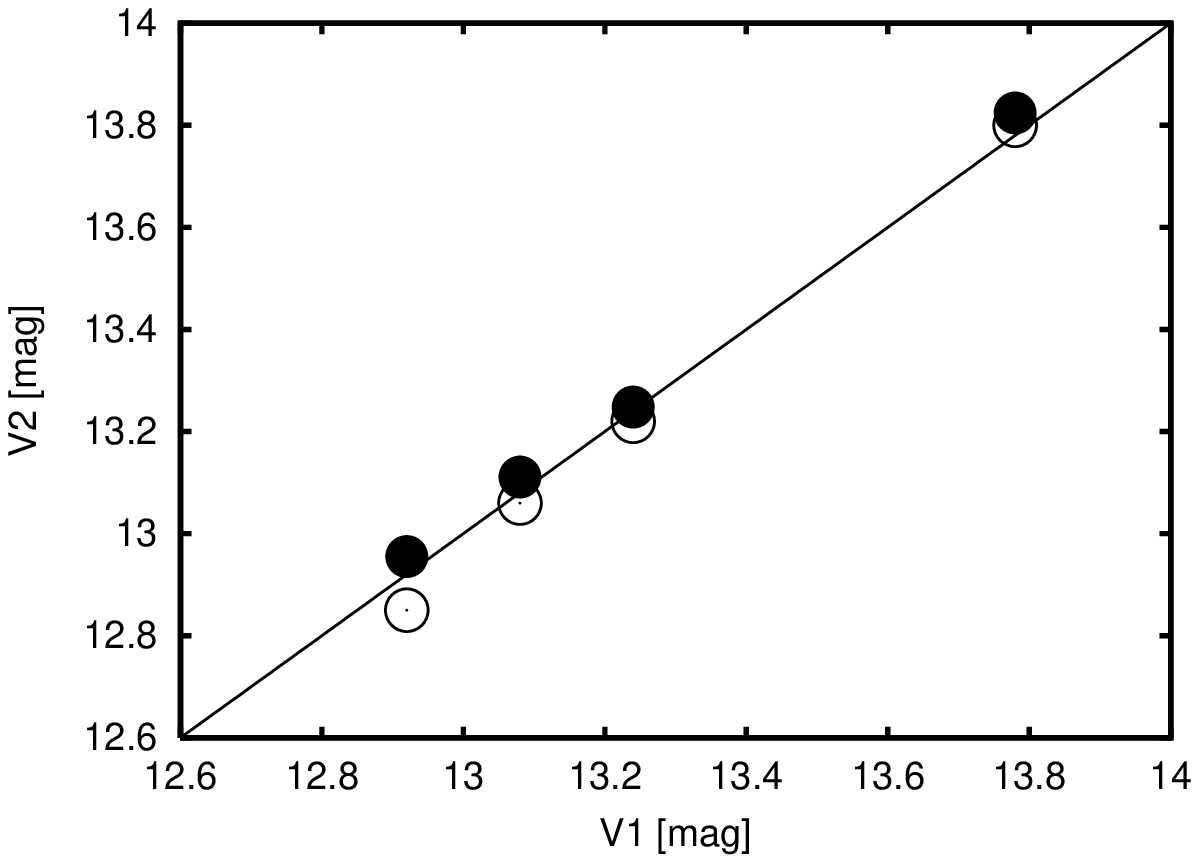,width=4.4cm}
\psfig{file=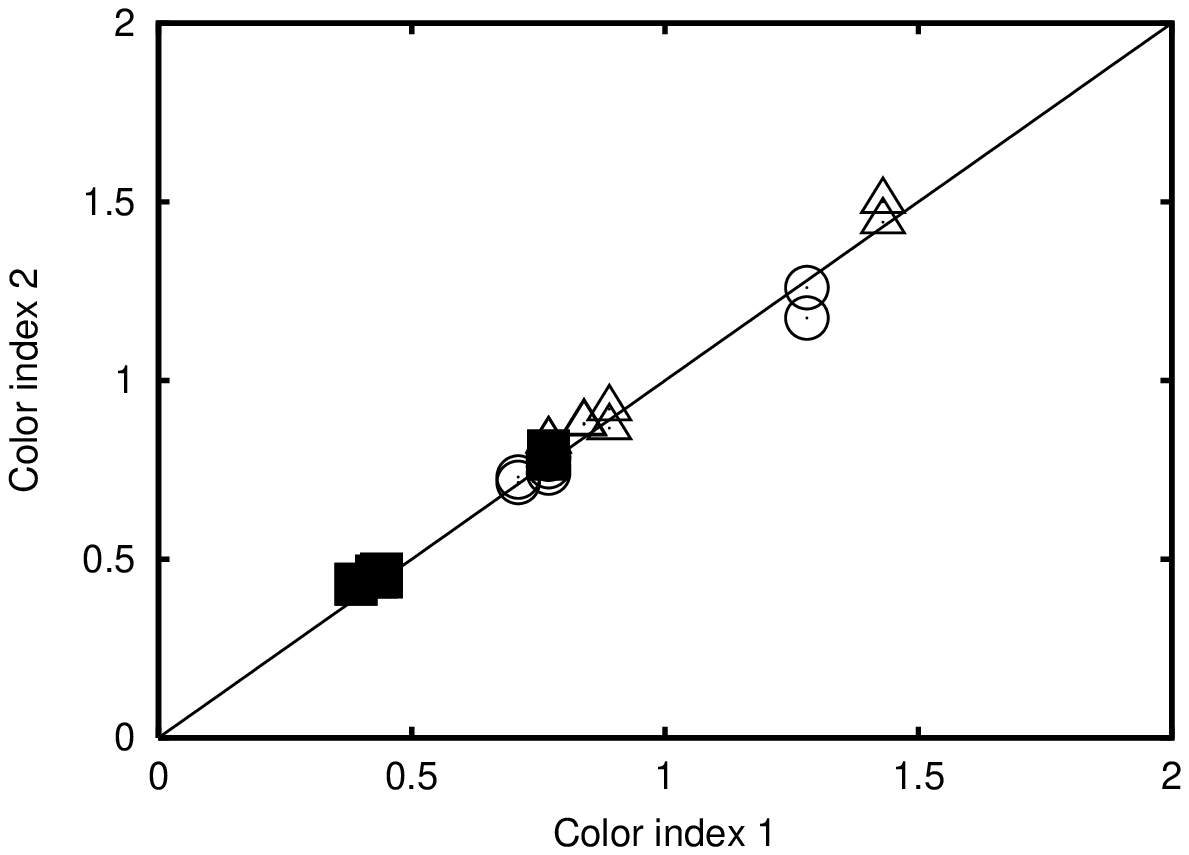,width=4.4cm}
\caption{Comparison of the standard sequences of Henden 
and Foley et al. plotted as a function of our standard magnitudes.
Left panel: $V$ magnitudes (open circles: Henden, 
filled circles: Foley et al.), 
right panel: $B-V$ (circles), $V-R$ (squares), $V-I$ (triangles).}
\end{center}
\end{figure}

\begin{figure*}
\begin{center}
\leavevmode
\psfig{file=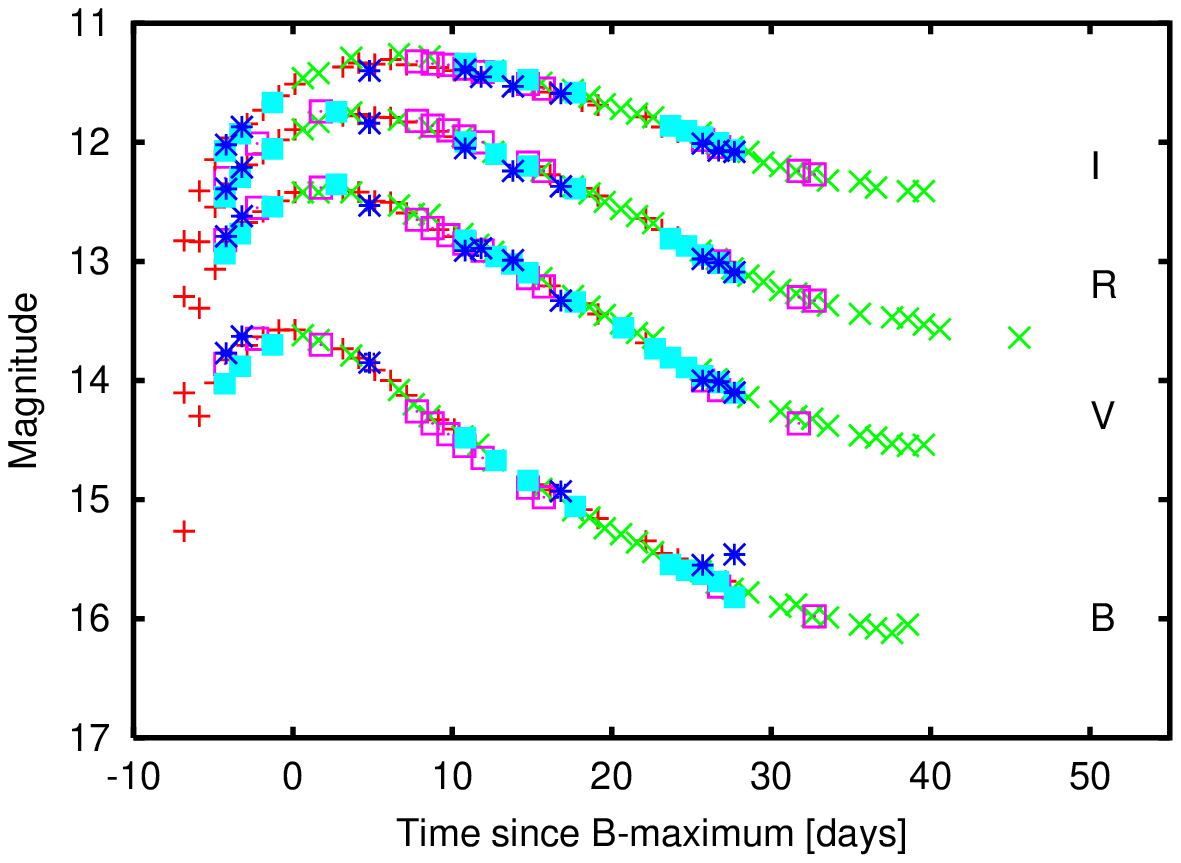,width=8.8cm}
\psfig{file=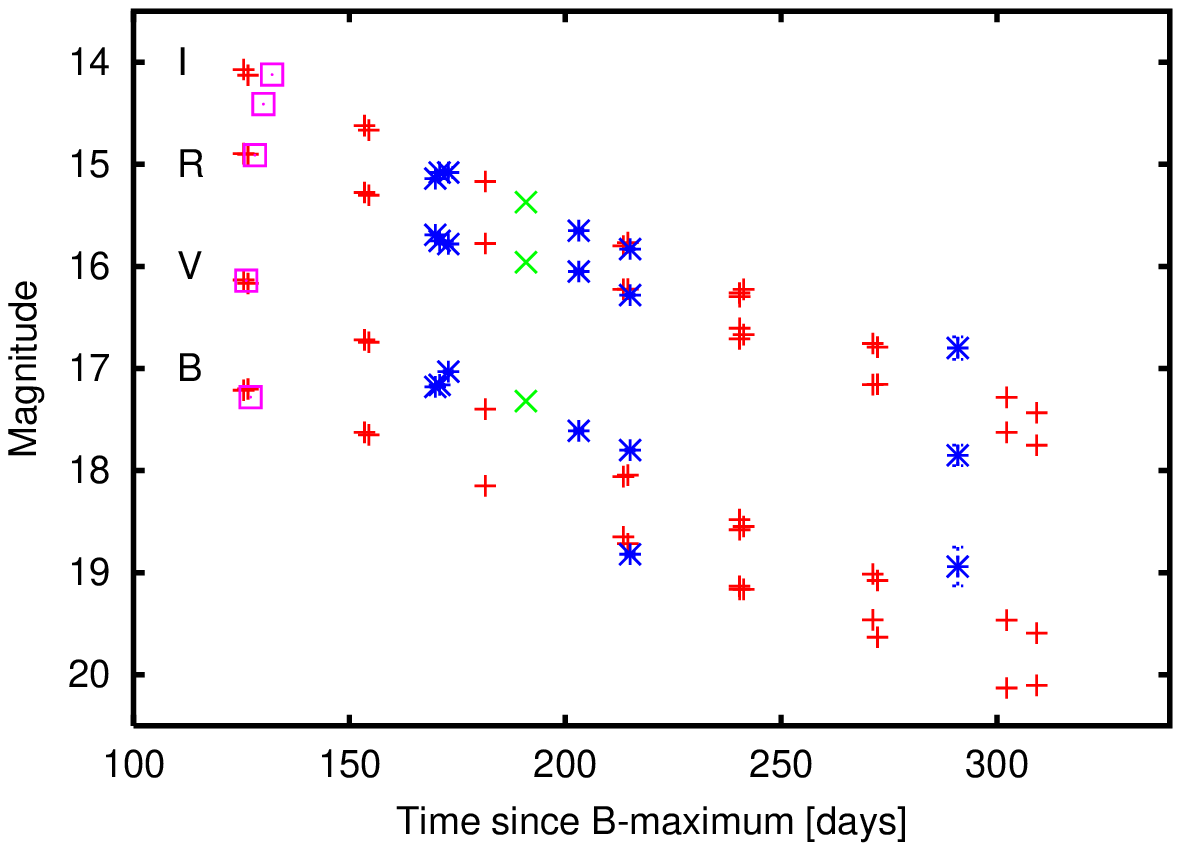,width=8.8cm}
\caption{The light variation of SN 2002ap around maximum (left panel)
and during the nebular phase (right panel). The data obtained with
different filters were shifted vertically for better visibility
(with $+0.5$, $0.0$, $-0.5$ and $-1.0$ mag for $B$, $V$, $R$ and $I$ data,
respectively). 
The symbols denote the following data sources: plus signs: 
Foley et al. (2003); crosses: Pandey et al. (2002); 
open squares: Yoshii et al. (2003); filled squares: Gal-Yam et al. (2002);
asterisks: present paper.}
\end{center}
\end{figure*}

The magnitudes listed in Table 1 were used to calculate the calibrated
standard magnitudes of SN~2002ap. The final results (after averaging
the magnitudes obtained from star A, B, C and D) are listed in Table~2.
Fig.3 shows the light curve in $BVRI$ bands.  In order to 
compare our photometry with other published data, we also plot
photometric data of SN~2002ap collected from literature. The agreement 
is satisfactory (note that the
tabulated light curve data in Foley et al. (2003) already 
contain the reddening correction - this was removed before
plotting in Fig.3). As usual, the phase is 
referenced to the moment of $B$-maximum, which was determined as 
JD 2452311.5 (note that it is 2 days less than the one given
by \cite{galy}).  

\begin{table*}
\caption{Photometry of SN~2002ap. The telescope codes are: A - 60 cm
Schmidt, Konkoly, B - 1 m RCC, Konkoly, C - 40 cm Cassegrain, Szeged.}
\begin{center}
\begin{tabular}{lccccc}
\hline
\hline
JD & $B$ & $V$ & $R$ & $I$ & Tel. \\
\hline
52307.30& 13.27(0.013)& 12.79(0.008)& 12.89(0.009)& 13.02(0.008) & A \\
52308.30& 13.13(0.015)& 12.62(0.006)& 12.71(0.008)& 12.87(0.007) & A \\
52316.30& 13.35(0.080)& 12.53(0.043)& 12.34(0.031)& 12.40(0.032) & C \\
52322.30& ...         & 12.91(0.041)& 12.55(0.022)& 12.39(0.024)  & C \\
52323.30& ...         & 12.89(0.067)& ...         & 12.45(0.035) & C \\
52325.30& ...         & 12.99(0.029)& 12.74(0.049)& 12.53(0.042) & C \\
52328.30& 14.43(0.051)& 13.33(0.046)& 12.87(0.052)& 12.59(0.028) & C \\
52337.20& 15.05(0.010)& 13.99(0.007)& 13.48(0.008)& 13.01(0.009) & A \\
52338.20& ...         & 14.01(0.014)& 13.51(0.008)& 13.07(0.015)  & A \\
52339.20& 14.96(0.029)& 14.10(0.016)& 13.59(0.021)& 13.08(0.019)  & A \\
52481.40& ...         & 17.18(0.007)& 16.19(0.008)& 16.14(0.006) & A \\
52482.40& ...         & 17.16(0.006)& 16.25(0.005)& 16.08(0.013)  & A \\
52484.40& ...         & 17.03(0.015)& 16.28(0.033)& 16.08(0.050) & A \\
52526.50& 18.32(0.022)& 17.80(0.015)& 16.78(0.008)& 16.83(0.012) & A \\
52514.60& ...         & 17.61(0.027)& 16.55(0.030)& 16.65(0.018) & B \\
52602.40& ...         & 18.94(0.190)& 18.35(0.110)& 17.80(0.125) & A \\
\hline
\end{tabular}
\end{center}
\end{table*}

\subsection{Constructing the bolometric light curve}

The bolometric light curve was constructed after combining all available
optical and near-IR measurements collected from the literature. After
dereddening with $E(B-V)=0.09$ mag (see next section), the magnitudes
were converted into monochromatic fluxes using the calibration of
\cite{bess}. The near-IR contribution was estimated from the $JHK$ data
of \cite{yosh}, while contribution from other bands were neglected.
Then, the bolometric fluxes were calculated by integrating the 
monochromatic fluxes with a simple trapezoidal rule. This resulted
in the ``observed'' bolometric flux as a function of time. Finally,
the absolute bolometric magnitudes have been determined applying the
true distance modulus $\mu_0 = 29.13$ mag derived in this paper
(see Sect.4). 

The absolute bolometric magnitudes have been compared with those
derived by \cite{yosh}. Because Yoshii et al. applied $\mu_0 = 29.5$
mag as distance modulus, their data were corrected to match the
slightly lower distance modulus used above. After correcting for
this difference, the two bolometric light curves showed very
good agreement. Thus, it is concluded that the various determinations
of the $UVOIR$ bolometric light curve of SN 2002ap have led to
consistent results, and the source of the main uncertainty 
of the bolometric light curve is the true distance modulus 
of the SN. 
 
The bolometric light curve of SNe can be described analytically
by the simple model of \cite{contardo}. In this model the
light curve has the form

\begin{equation}
m_{bol} = { m_0 + \gamma (t-t_0) + g_0 \exp[-{(t-t_0)^2 / (2 \sigma_0^2)}]
\over {1 - \exp [{ {t_2-t} \over \sigma_2} ]} }
\end{equation}  

Note that we applied only one gaussian component, because the
bolometric light curve of SN 2002ap does not show significant
bump on the descending branch. The light curve parameters 
were determined via least-squares fitting as 
$m_0 = 13.82 \pm 0.1$, $t_0 = 310.0 \pm 0.5$, $t_2 = 297.95 \pm 0.02$,
$\gamma = 0.0172 \pm 0.0005$, $\sigma_0 = 18.3658 \pm 0.0004$, 
$\sigma_2 = 3.1677 \pm 0.0002$, $g_0 = 1.4684 \pm 0.2$. 
Among these, the most interesting parameter is the late-time 
decline rate $\gamma$ that is physically linked to the
Co-Fe decay and the radiation transport in the expanding SN ejecta.
The value derived above, $\gamma = 0.017$ mag/day 
is in good agreement with
the one published by \cite{pande} ($0.0199 \pm 0.0004$).   

In Fig.4 we plot the ``observed'' bolometric light curve 
($m_{bol} = M_{bol} + \mu_0$) together with
the analytic model given above and the expected decline 
rates due to the Ni-Co-Fe radioactive decay. As seen in Fig.4, 
the decline of SN 2002ap is significantly faster than the Co-Fe 
radioactive decline rate ($\tau = 0.0098$ mag/day), 
as also noted by \cite{pande} and \cite{yosh}. They
explained this as a leakage of $\gamma-$photons from the atmosphere,
indicating a transparent, less massive ejecta. Our new late-time 
photometry fully supports these conclusions. The connection between
the physical state of the expanding atmosphere and the bolometric
light curve is analyzed further in Section 4. 

\begin{figure}
\begin{center}
\psfig{file=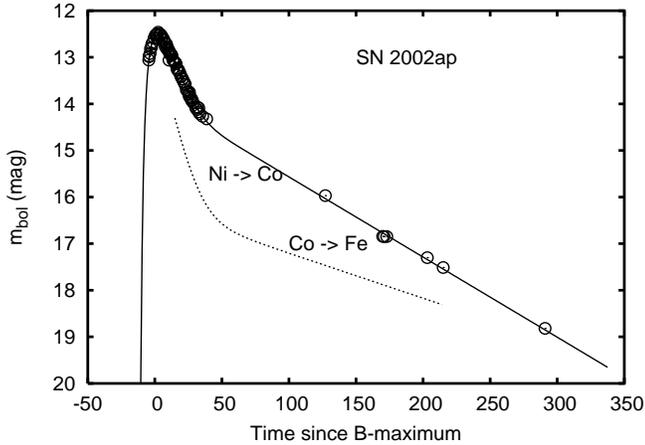,width=8.8cm}
\caption{The bolometric light curve of SN 2002ap. The continuous line
represents the fitted analytic function given by Eq.1, while the dotted line
shows the rate of the energy input by the indicated radioactive decay.}
\end{center}
\end{figure}

\subsection{Optical spectroscopy}
The spectroscopic observations were carried out at David Dunlap Observatory (DDO)
between the 2nd and 25th February, 2002 ($t$ = $-3$ to $+20$ days), 
with the Cassegrain spectrograph attached
to the 74" telescope. Two different setups have been applied: the 100 lines/mm 
grating was used for the low dispersion observations, while the 1800 lines/mm
grating was selected for obtaining high resolution spectra. 
The reciprocal dispersion was 3.60 \AA /pixel and 0.2 \AA /pixel, respectively.
The low-resolution spectra extended from 4000 to 7500 \AA, while the
high-resolution spectrum was centered on Na D and covered a 200 \AA-wide
spectral interval. The slit width was 306 microns (corresponding to 1.8 arcsec
on the sky) for both observing modes. The seeing was between 2 - 3 arcsec
during the observations.
 
The reduction was done in the standard way using {\it IRAF}. 
Flux calibration was performed via the spectra of comparison
stars HD 26793 (spectral type B9V) and HD 74280 (B3V), obtained before
or after the SN observation on the same night with the same setup. 
The slit of the DDO spectrograph has fixed E-W direction,
so the observations could not be taken at the parallactic angle. 
During the observations the airmass of the SN was always less than 2,
but differential light loss may have reduced the observed fluxes in the
blue part of the spectra (especially below 4500 \AA). 
However, this effect is probably not strong, because the shapes of our 
spectra are similar to those collected from literature (see below).
 
After the calibrations, the spectra were de-reddened using $E(B-V)$ = 0.09
mag (discussed below) and corrected for the redshift of the
host galaxy ($z$ = 0.0021, \cite{foley}). Table 3 contains the journal 
of spectroscopic observations.  

\begin{table}
\caption{Journal of spectroscopic observations. The coloumns contain the
following data: date and JD of the observation, phase relative to 
B-maximum (JD 2452311.5, see text), central wavelength (in \AA),
spectral resolution, airmass of the SN and the parallactic angle
at the time of the observation (in degrees).}
\begin{center}
\begin{tabular}{ccccccc}
\hline
\hline
Date & JD-2452000 & Phase & $\lambda_0$ & $\lambda / \Delta \lambda$ & airmass & PA\\
\hline
2002.02.02 & 308.49 & $-3$ & 5800 & 800   &  1.22 & 31\\
2002.02.03 & 309.50 & $-2$ & 5800 & 800   &  1.25 & 34\\
2002.02.06 & 312.54 & $+1$ & 5890 & 14700 &  1.46 & 44\\
2002.02.09 & 315.54 & $+4$ & 5800 & 800   &  1.60 & 46\\
2002.02.15 & 320.53 & $+9$ & 5800 & 800   &  1.54 & 46\\
2002.02.18 & 323.50 & $+12$ & 5800 & 800   &  1.45 & 44\\
2002.02.24 & 330.49 & $+19$ & 5800 & 800   &  1.60 & 45\\
2002.02.25 & 331.52 & $+20$ & 5800 & 800   &  1.82 & 47\\
\hline
\end{tabular}
\end{center}
\end{table}

\begin{figure*}
\begin{center}
\psfig{file=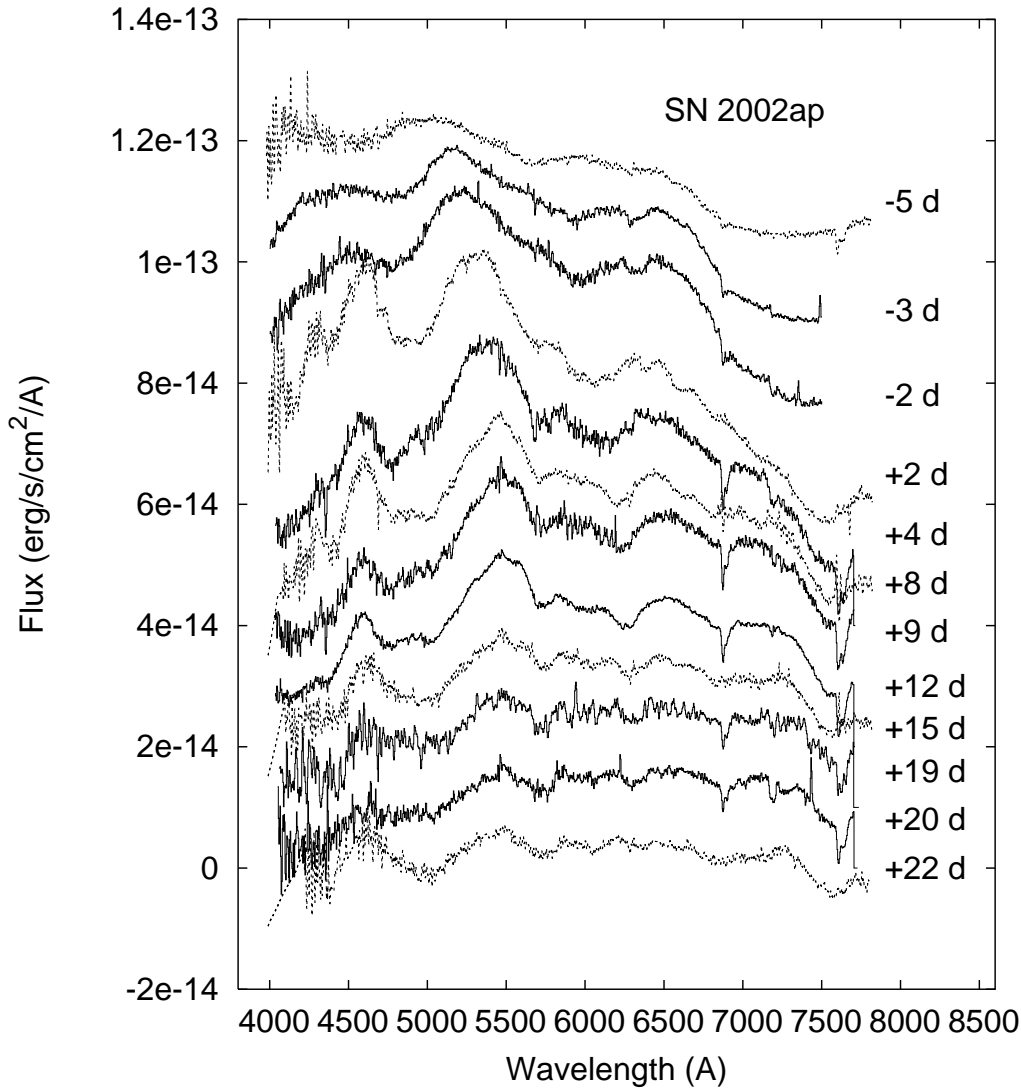,width=14cm}
\end{center}
\caption{The spectral evolution of SN~2002ap around maximum 
(solid line: DDO spectra, dotted line: Wise Obs. spectra).
The phase relative to $B$-maximum is indicated on the right-hand
side. Each spectrum was shifted vertically for better visibility.}
\end{figure*}

The final, redshift-corrected and dereddened spectra are plotted 
in Fig.5, together with the spectra taken by \cite{galy} 
at Wise Observatory. 
The characteristic features of hypernovae are apparent: very
blue, almost featureless spectrum at early phases, very 
broad, strong troughs around maximum light, and flat 
continuum with shallow, broad absorption lines at
the end of the photospheric phase. These spectra are
very similar to those presented by \cite{b8} and
\cite{foley}. Kinugasa et al. showed that the deep
absorption troughs developing around maximum are
due to \ion{Ca}{ii}, \ion{Fe}{ii}, \ion{Si}{ii}, 
\ion{Ni}{ii}, \ion{Co}{ii}, \ion{O}{i} and \ion{Na}{i}.

Radial velocities were determined from the usual \ion{Si}{ii}
$\lambda$6355 line. Table 4 lists the derived velocities, 
together with other available data collected from the literature.
The uncertainties were estimated from the width of the
\ion{Si}{ii} absorption trough. These errors are rather large 
due to the strong broadening at the very early phases and the decreasing
line strength at late phases. 
Fig.6 contains the comparison of \ion{Si}{ii} velocities 
of hypernovae SN~1998bw (\cite{patat}) and SN~1997ef
(\cite{maz97}). This figure agrees very well with a similar
plot presented by \cite{b8}, and supports their
conclusion that SN~2002ap evolved faster than the
other two well-observed hypernovae (see also 
\cite{foley}). The radial velocities will be used 
in Sect.3 and 4.

\begin{figure}
\begin{center}
\psfig{file=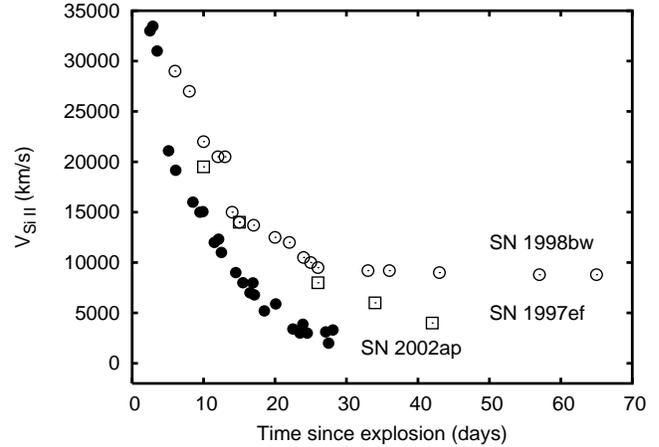,width=8.8cm}
\end{center}
\caption{The velocity curve of hypernovae SN~2002ap (filled circles),
SN~1998bw (open circles, \cite{patat}) and SN~1997ef 
(open squares, \cite{maz97}).}
\end{figure}

\begin{table}
\caption{Radial velocities of SN~2002ap inferred from
the Doppler-shift of the \ion{Si}{ii} $\lambda$6355 line. 
The second column gives the phase with respect to
the date of explosion (JD 2452303.4, \cite{mazz}).
The errors of the radial velocities (fourth coloumn)
were estimated from the width of the \ion{Si}{ii} line.
The meaning of the reference codes is: K - \cite{kinu}; 
W - \cite{galy}; pp - present paper.}
\begin{center}
\begin{tabular}{ccccc}
\hline
\hline
JD & $t$ & $v_r$(Si II) & $\sigma$ & ref. \\
(2452000+) & (days) & (kms$^{-1}$) & (kms$^{-1}$) &  \\
\hline
305.9 & 2.5 & 33000 & $\pm 3000$ & K \\
306.3 & 2.9 & 33465 & $\pm 3000$ & W \\
306.9 & 3.5 & 31000 & $\pm 2000$ & K \\
308.5 & 5.1 & 21097 & $\pm 2000$ & pp \\
309.5 & 6.1 & 19162 & $\pm 2000$ & pp \\ 
311.9 & 8.5 & 16000 & $\pm 1500$ & K \\
312.9 & 9.5 & 15000 & $\pm 1500$ & K \\
313.3 & 9.9 & 15055 & $\pm 1000$ & W \\
314.9 & 11.5 & 12000 & $\pm 1500$ & K \\
315.5 & 12.1 & 12317 & $\pm 1000$ & pp \\
315.9 & 12.5 & 11000 & $\pm 1500$ & K \\
317.9 & 14.5 & 9000 & $\pm 1500$ & K \\
318.9 & 15.5 & 8000 & $\pm 1500$ & K \\
319.9 & 16.5 & 7000 & $\pm 1500$ & K \\
320.3 & 16.9 & 7974 & $\pm 1500$ & W \\ 
320.5 & 17.1 & 6794 & $\pm 1500$ & pp \\
321.9 & 18.5 & 5200 & $\pm 1500$ & K \\
323.5 & 20.1 & 5897 & $\pm 1500$ & pp \\
325.9 & 22.5 & 3400 & $\pm 2000$ & K \\
326.9 & 23.5 & 3000 & $\pm 2000$ & K \\
327.3 & 23.9 & 3867 & $\pm 2000$ & W \\
327.9 & 24.5 & 3000 & $\pm 2500$ & K \\
330.5 & 27.1 & 3112 & $\pm 2500$ & pp \\
330.9 & 27.5 & 2000 & $\pm 2500$ & K \\
331.5 & 28.1 & 3301 & $\pm 2500$ & pp \\                 
\hline
\end{tabular}
\end{center}
\end{table}

A single high-resolution spectrum was obtained on 6th Feb. 2002
centered around Na D (Fig.7). The presence of the narrow 
dublet at zero velocity is clearly indicative of interstellar
absorption originating from Milky Way ISM. The equivalent
width of the two components was measured to be $0.30 \pm 0.04$ 
and $0.31 \pm 0.05$ \AA.
According to \cite{barbon}, the total EW of the dublet 
corresponds to $E(B-V) ~=~ 0.15 \pm 0.02$ mag. The more
recent relation of \cite{muzw} (that uses only the
EW of the $\lambda 5891$ \AA~component) results in 
$0.12 \pm 0.01$ mag. The uncertainties given here are
formal errors of the EW measurement errors. A more
realistic error bar that reflects the uncertainties
of the applied empirical relations is in the order
of $\pm 0.1$ mag. Despite of the lower S/N of our high-resolution
spectrum, these estimates agree quite well with the
value derived by \cite{foley} ($E(B-V) \approx 0.1$).
This confirms the conclusion of many previous papers that the
reddening of SN~2002ap is low, but not negligible. 

For such low reddening, the all-sky map of 
\cite{sfd} usually works better. This gives 
$E(B-V)_{gal} = 0.071$ mag as the galactic component
(that is also probed by the observed zero velocity NaD dublet).
This is somewhat lower than the values listed above,
but it is within the global uncertainty of the above 
estimates. The contribution of the host galaxy ISM was measured
by \cite{taka}, and found to be as low as 0.02 mag. The
sum of these two components gives $E(B-V) ~=~ 0.09$ mag,
which was used in almost all previous papers on SN~2002ap.
Therefore, we have adopted $E(B-V) ~=~ 0.09 \pm 0.02$ mag
(estimated uncertainty) throughout the rest of this work. 
Our spectroscopic result, although with higher uncertainty,
supports this low reddening value. 
 
\begin{figure}
\begin{center}
\psfig{file=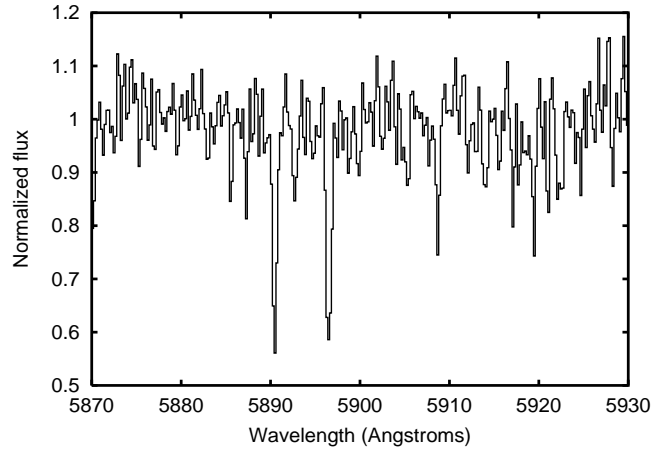,width=8.8cm}
\end{center}
\caption{Na D lines in the spectrum of of SN 2002ap.}
\end{figure}

\section{Distance estimate}

Distance is one of the most important parameter for SNe,
because the physical parameters, such as luminosity and
kinetic energy, can be inferred reasonably only if the
distance is known. Core-collapse events, such as hypernovae,
usually exhibit large variation in their peak luminosity and ejected
mass (\cite{hamu2}), thus, distance determination is necessary to infer
these parameters from the observed ones.

The distance of a hypernova is not an easy parameter to determine.
Because these events are far from being ``standard candles'',
a pure photometric distance measurement is not possible at present.
A possible solution for this problem may be the determination
of the distance of the host galaxy itself, via another,
independent method. This might work at least for the nearby
galaxies, such as M74. Surprisingly, such pre-determined
distances for M74 are quite uncertain (see Section 1).
Thus, in order to obtain reliable parameters, 
a distance measurement of SN~2002ap itself would be very useful.

\subsection{The Expanding Photosphere Method}

We have applied the Expanding Photosphere Method (EPM) 
for measuring the distance to SN~2002ap. This method was developed
by \cite{kirshner}, and was used many times for measuring
distances of Type II SNe (e.g. \cite{schmidt}; \cite{hamuy}). 
The main assumptions of this method are as follows:
\begin{itemize}
\item{the expansion is spherically symmetric and homologous}
\item{there is a well-defined photosphere in the ejecta, below which the gas
is optically thick}
\item{the photosphere radiates as a (diluted) blackbody 
(parametrized by the dilution factor $\zeta$).}
\end{itemize}

Because Type II SNe eject a massive H-rich envelope, they
have optically thick atmospheres, in which the location of
the photosphere is determined by the position of the 
H recombination front. This phase lasts for months, therefore
the EPM can be applied for data obtained over a long time interval.

It is very probable that none of the above assumptions are 
fully applicable for SN~2002ap, which was very different
from a Type II SN. According to the spectroscopic observations
(including those presented in this paper), the photosphere 
disappeared quickly, and after a few weeks the SN entered the
nebular phase. Therefore, the EPM can be applied only for
a few data points just before and after maximum light. 
Moreover, the atmosphere of 
hypernovae could be very different from that of a Type II SN,
and the emergent radiation may be far from being Planckian.
In their very thorough study of Type II SN atmospheres, 
\cite{eastman} showed that the dilution factor $\zeta$ for a 
particular atmosphere is usually between 0.4 - 1.5 depending on
the effective temperature, density profile and the choice
of the wavelength regime for determining the colour temperature.
Because $\zeta$ directly influences the EPM-based distances,
neglecting the deviation from Planckian flux can lead to a
40\% systematic error in the derived distance. 
For hypernovae, such kind of problems can be studied only via 
detailed model atmospheres, which is beyond the
scope of this paper. Nevertheless, we assumed that the EPM
can be applied to SN~2002ap, as a first approximation,
until the more sophisticated model atmosphere computations
are available.

Perhaps the weakest point of the feasibility of the EPM is
the assumption of spherical symmetric expansion. According
to the polarization results (Section 1), there are 
indications that the expansion of SN~2002ap was definitely
asymmetric at early phases. The measured net polarization
(about 2 \%) may suggest an asymmetry of about 10 \%
(\cite{kawa}), although the polarization angle varied in
time, which means that the asymmetry, if indeed present, may
be of complex nature. Another possibility may be the 
presence of higher interstellar polarization than Kawabata et al.
estimated. In this case the measured polarization is not due
to the elongated shape of the ejecta. Indeed, recently \cite{leo2}
has discovered evidence for extremely high polarization efficiency
in NGC 3184, where a high degree of polarization was measured for
SN 1999gi that otherwise was only slightly reddened.
It is also worth noting that 
model computations for SN~2002ap could reproduce the early part of the
light curve using sperical as well as non-spherical models
(see e.g. \cite{maeda} for details). In the followings we 
assume, for simplicity, that the expansion of SN~2002ap was 
spherically symmetric during the photospheric epoch, 
but note that deviations from sphericity may result in 
systematic errors in the distance determination.  
Discussion of other possible sources of random and
systematic uncertainty is presented at the end of this Section.

The formulae of the EPM are straightforward. According to
the homologous expansion, the radius of the photosphere can
be expressed as:
\begin{equation}
R = v_{phot} (t - t_0) + R_0
\end{equation}
where $t_0$ is the moment of explosion. $t_0$ is a crucial
parameter for the distance determination. The angular size
of the photosphere, $\theta = R / D$, where $D$ is the
distance, can be expressed as (\cite{kirshner})
\begin{equation}
\theta = \sqrt{ {f_\lambda} \over {\zeta_\lambda^2 \pi B_\lambda (T_\lambda)} }
\end{equation}
where $f_\lambda$ is the observed flux density, $ B_\lambda (T_\lambda)$ 
is the Planck function and $\zeta_\lambda$ is the dilution factor.

In this paper we used the $UVOIR$ bolometric light curve of SN~2002ap
(Section 2) for estimating the surface brightness - temperature relation. 
The dilution factor is assumed to be $\zeta = 1$ (as a first approximation) 
and Eq.3 can be written as
\begin{equation}
\theta = \sqrt{ {f_{bol}} \over {\sigma T_{eff}^4} }
\end{equation}
where $f_{bol}$ is the observed bolometric flux density (in erg/s/cm$^2$) 
and $\sigma$ is the Stefan-Boltzmann constant. $f_{bol}$ is given from the
bolometric light curve discussed in Sect.2. The effective temperature
was estimated from hypernova model atmospheres.

We used the hypernova model atmospheres by \cite{maz97} constructed for 
SN~1997ef. For several epochs, they tabulated the effective temperature
and the calculated $B - V$ color of the photosphere. We fitted a simple
linear function to these data and found 
\begin{equation}
T_{eff} ~=~  -0.122(\pm 0.03)~(B-V) ~+~ 3.875(\pm 0.03). 
\end{equation}
Combining Eq.4 and 5, the angular size of the photosphere $\theta$ can
be derived from the observations. 

The photospheric velocities were assumed to be equal to
the radial velocities based on the usual \ion{Si}{ii} line (Sect.2).
The hypernova models of \cite{maz97} give some support to our assumption,
as their computed \ion{Si}{ii} velocities are similar to the 
photospheric ones.

\begin{figure}
\begin{center}
\leavevmode
\psfig{file=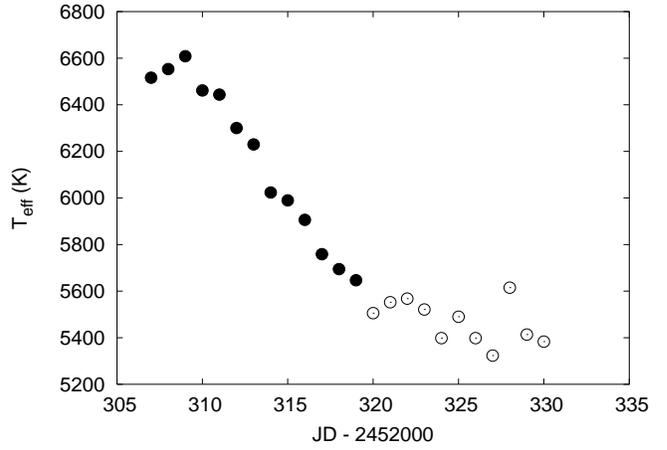,width=8.8cm}
\caption{The effective temperature of SN 2002ap (via Eq.5) as a function of time.
The filled circles correspond to data obtained during the expanding photosphere
phase (see Fig.9).}
\end{center}
\end{figure}

\begin{figure}
\begin{center}
\leavevmode
\psfig{file=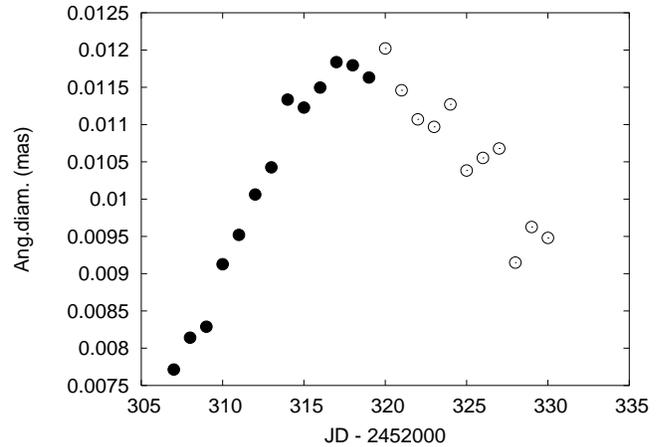,width=8.8cm}
\caption{The variation of the angular size (in mas) of the photosphere of SN 2002ap.
The expansion of the photosphere lasted only about 12 - 13 days after explosion.}
\end{center}
\end{figure}

Having the angular size and the velocity determined from the observations
at several epochs, the distance can be derived via Eq.2 and 4. Following
\cite{schmidt}, their combination yields
\begin{equation}
t ~=~  D~ k ~( {\theta \over v_{phot}} ) ~+~ t_0 ,
\end{equation}  
where $D$ is in Mpc, $v_{phot}$ is in kms$^{-1}$, $t$ is in days 
and $k = 3.5697 \cdot 10^{14}$ is the conversion factor between the
quantities. Plotting $t$ versus $\theta / v_{phot}$, $D$ and 
$t_0$ can be determined by linear regression. 

\begin{figure}
\begin{center}
\leavevmode
\psfig{file=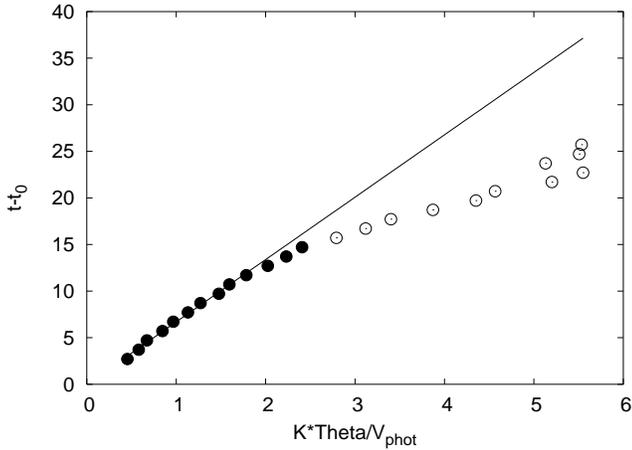,width=8.8cm}
\caption{The fit of Eq.6 to the data of SN~2002ap, yielding the distance and
the moment of explosion.}
\end{center}
\end{figure}

The results for SN~2002ap are plotted in Figs.8, 9 and 10. Fig.8 contains
the plot of the temperature variation (from dereddened $(B-V)_0$ data, via
Eq.5). Fig.9 shows the angular size calculated from Eq.4. It is apparent that
the expansion of the photosphere stopped at about JD 2452317, so the
expansion phase did not last long. Note, that, as the referee pointed 
out, this does not mean that the {\it photospheric} phase has ended, 
because the spectra clearly indicate that the SN is still in 
the photospheric phase (see Fig.5). The EPM was applied only for data 
obtained before this epoch. 
The fitting of Eq.6 to the combined data is
presented in Fig.10. The linear regression resulted in $D$ = 6.7 $\pm 0.5$ 
Mpc and $t_0$ = 2452304.3 $\pm 1.0$ as the distance and the moment of
explosion, respectively. It is clear from Fig.10 that the relation becomes
non-linear when the expansion phase ends, thus, at least one of the
assumptions of the EPM became invalid at that time. The quoted uncertainties
are formal (random) errors due to the fitting procedure. 

These results show remarkably good agreement with previous ones, 
determined via other methods. The EPM-distance is almost perfectly equal
to the photometric distance of M74 when the \cite{sfd}
reddening correction is applied ($d = 6.8$ Mpc, see Sect.1).
The moment of explosion agrees within the uncertainties 
with the one derived by \cite{mazz} ($t_0 = 2452303.4$, Jan. 28, 2002)
from model computations. It also worths noting that the derived
explosion date is only 1 day prior to the discovery of 
SN~2002ap (see Sect.1).

\subsection{Uncertainties of the EPM-distance}
As mentioned above, difference between the assumptions of the EPM and
the true physical state of the SN atmosphere may lead to significant
uncertainty in the inferred distance. Despite the good agreement
between the new EPM-distance and other distance estimates, all 
these may suffer from quite large systematic errors. Recently,
\cite{leo3} presented a thorough test of the EPM- and Cepheid-based
distances of SN~1999em and NGC~1637. They concluded that the 
Cepheid-based distance is 50 \% greater than the EPM-based distance,
which means that there is serious disagreement between these two
measurement methods. They also showed that, at present, the 
Cepheid-based distance is more favourable than the EPM-based one,
thus, the EPM-distances are plagued by at least
50\% systematic error. This leads to a 60 - 70 \% total 
(systematic plus random) uncertainty for the EPM. 

What are the possible sources of such high uncertainty?
\cite{leo3} mention the dilution factor and the radial
velocity measurement technique as the most probable ones.
The need for a precise dilution factor based on model atmosphere
tuned for the particular SN has been emphasized by \cite{eastman}. 
However, their dilution factors may also suffer from
uncertainies, because independent calculations by other groups
have lead to significantly different factors (see \cite{leo3}
for extensive discussion and references). These problems
are probably more enhanced for the present study, due to 
the atmospheric properties of hypernovae. Since there was
no detailed model atmosphere for SN~2002ap at our disposal,
we had to assume $\zeta = 1$, which may lead up to 50 \%
systematic error in the distance (as it was mentioned in 
Section 3.1). This error can be even larger if the atmosphere
of SN~2002ap were very different from a Type II SN atmosphere 
so that the dilution factor was quite different from those
of \cite{eastman}. On the other hand, the application 
of a hypernova model of \cite{maz97} (constructed
for SN~1997ef) may somewhat reduce this large uncertainty.
As it is explained above, we assigned the {\it model} effective
temperature (via Eq.5, by comparing the observed and 
predicted $B-V$ color indices) to the {\it bolometric} 
flux data. This technique may have the advantage of using
the $UVOIR$ ``bolometric'' fluxes that are integrated over
a long wavelength regime, thus, it is not sensitive to the
chosen wavelength and wavelength-dependent 
dilution factors. On the other hand, the uncertainties 
of the derived bolometric fluxes are probably in the 
same order of magnitude as that of the dilution factor.

The systematic errors caused by the measurement method of
radial velocities are also quite large for SN~2002ap
because of the very broad lines and strong blends in the 
spectra, especially at the early phases. 
Therefore, referring to \cite{leo3}, we estimate the 
systematic uncertainty of the EPM-based distance as 
60 \% that means $\pm 4$ Mpc for SN~2002ap. The total
(random plus systematic) error of the distance 
becomes $\pm 4.5$ Mpc, or about 70 \%. 

The consistency between the
EPM-distance and the other photometric distances of M74 
may support the applicability of the EPM for hypernovae.
However, it is important to note that \cite{leo3} 
found similar agreement between the EPM-distance and
the ``brightest red supergiants'' (BRSG) distance
(\cite{sohn2}) for NGC~1637. Among the distance measurement
methods discussed by \cite{leo3}, the BRSG technique
is the only one that gives consistent result with
the EPM. Therefore, this agreement does not, unfortunately,
reduce the large systematic uncertainty of the EPM-distance
for M74. 
 
We conclude that our first attempt to apply the EPM-technique
to a hypernova resulted in distance estimate for SN~2002ap
that has remarkably good internal precision (good linearity,
low scattering in Fig.10.) On the other hand,
the systematic uncertainty of the EPM-distance is quite 
high due to the problematic issue of the dilution factors
as well as the complexity of hypernova spectra at 
the early phases.  Note that
the EPM-distance is 0.6 Mpc lower than the one used in previous
papers (7.3 Mpc), thus, the inferred physical parameters of
SN 2002ap may need some revision, if we adopt this new 
distance. For example, the nickel mass may be somewhat lower
than derived previously (e.g. \cite{mazz}; \cite{pande};
\cite{foley}).

\section{Model computations and physical parameters}
 
\begin{table*}
\begin{center}
\caption{Computed models of SN~2002ap. The meaning of the symbols are:
$M$: ejected mass (in $M_{\odot}$); $M_{Ni}$: mass of synthesized Ni (in $M_{\odot}$); 
$x_0$: fractional core radius; $n$: power-law exponent of density; 
$v_{exp}$: expansion velocity (in (kms$^{-1}$);
$\kappa_{\gamma}$: $\gamma$-ray opacity (in cm$^2$g$^{-1}$); 
$\kappa_{e^+}$: positron opacity; $\kappa_{opt}$: opacity for optical photons;
$E_{51}$: kinetic energy of the explosion (in $10^{51}$ ergs) 
The asterisk indicates that the parameter was kept fixed during the fitting.}
\begin{tabular}{lccccccccc}
\hline
\hline
Model&$M(M_{\odot})$&$M_{Ni}(M_{\odot})$&$x_0$&$n$&$v_{exp}$&$\kappa_{\gamma}$
&$\kappa_{e^+}$&$\kappa_{opt}$&$E_{51}$\\
\hline
A & 1.0 & 0.07 & 1.0* & 0.0* & 35000 & 0.1 & 5.0 & 0.009 & 7.35\\
B & 1.0 & 0.04 & 0.01* & 2.1 & 35000 & 0.005 & 0.1 & 0.03 & 3.84 \\
C1 & 1.0 & 0.07 & 0.15* & 2.1 & 40000 & 0.027* & 1.4 & 0.025 & 5.67 \\
C2 & 2.0* & 0.045 & 0.15* & 2.1 & 35000 & 0.027* & 1.0 & 0.012 & 8.69 \\
\hline
\end{tabular}
\end{center}
\end{table*}

Some basic physical parameters of the SN explosions can be 
estimated by comparing the observations with model computations of
the light curves and/or spectra. This usually requires sophisticated
calculations including hydrodynamics, radiative transfer and atomic
physics, which are beyond the scope of this paper. Instead, 
in the followings we use a simple analytic (``toy''-) model to
estimate the fundamental parameters of SN 2002ap, such as Ni-mass
and kinetic energy of the ejecta.
 
The bolometric light variation of SNe can be qualitatively described by simple
analytic models (\cite{arnett1}; \cite{arnett2}; \cite{iwam}; \cite{maeda}), 
assuming homologous expansion of the ejected envelope
and (probably over-)simplified deposition of gamma-rays. 
It is known that SNe are powered by radioactive decay of 
$^{56}$Ni $\rightarrow$ $^{56}$Co $\rightarrow$ $^{56}$Fe.
The basic parameter characterising the
emitted energy is $M_{Ni}$, the mass of $^{56}$Ni synthesized 
during the explosion. The gamma-rays originating from
radioactive decay are deposited and thermalized in the ejecta via
Compton-scattering. Because the ejecta is optically thick at early
phases, the timescale of the radiative transfer of photons to 
the surface is comparable to the timescale of the expansion (characterized
by $v_{exp}$, the velocity of the outmost part of the ejecta), thus, the
energy remains trapped in the optically thick atmosphere. As the
ejecta expands, the density and the opacity decreases, and the deposited
energy can escape more and more quickly. Thus, the light curve has a 
peak at 15-20 days past explosion, when
the emitted luminosity becomes roughly equal to the instantaneous 
rate of the energy deposition (\cite{arnett2}). Because the transparency
of the ejecta quickly decreases with time, the light variation after maximum
mimics the radioctive decay law and the physics of gamma-ray deposition.

The very simple light curve model used in this paper is based on the
assumption of free, homologous expansion of the ejecta. The ejecta has
a radius $R(t)$ that increases with time as
\begin{equation}
R(t) = R_0 + v_{exp} t 
\end{equation}
where $R_0$ is the (negligible) initial radius of the progenitor and
$v_{exp}$ is the (constant) expansion velocity. 
The velocity of a thin shell at fractional radius 
$x = r(t) / R(t)$ has a velocity
\begin{equation}
 v(x) = x v_{exp}.
\end{equation}
 
The density structure of the ejecta is assumed to be consisting of
an inner core with constant density, and an outer shell where the density
is decreasing as power-law:
\begin{equation}
\rho(x) = \rho_0 (x_0 / x )^n  
\end{equation}
where $x_0$ is the fractional radius of the core, 
$\rho_0$ is the density in the core (i.e. for $x < x_0$) and $n$ is the
power-law exponent. As a consequence of the expansion, $\rho_0$ decreases
in time as 
\begin{equation}
\rho_0 (t) = {M \over {4 \pi f(x_0)}} ~ {1 \over R(t)^3} 
\end{equation}
where $f(x_0) = ( {x_0}^3 / 3 + {x_0}^n (1- {x_0}^{3-n})/(3-n) )$
is a geometric factor due to the assumed density structure, and
$M$ is the total mass of the ejecta.  
With these assumptions the kinetic energy can be expressed as
\begin{equation}
E_k = {3 \over 10} ~ {{3-n} \over {5-n}} ~
{{5 x_0^n - n x_0^5} \over {3 x_0^n - nx_0^3}} ~ M v_{exp}^2
\end{equation}

The energy production released as gamma-rays is described by the 
exponental decay law:
\begin{eqnarray}
 E_{\gamma} & = & M_{Ni} ~ (S_{Ni} + 0.82 ~ S_{Co}) \\
 S_{Ni} & = & E_{Ni}~ \exp(- \lambda_{Ni} t) \\
 S_{Co} & = & E_{Co}~ ( \exp(- \lambda_{Co} t) - \exp (- \lambda_{Ni} t) )
\end{eqnarray} 
where $E_{Ni} = 3.70 \cdot 10^{10}$ erg/g, $E_{Co} = 6.76 \cdot 10^9$ erg/g, 
$\lambda_{Ni} = 1.3152 \cdot 10^{-6}$ s$^{-1}$, 
$\lambda_{Co} = 1.0325 \cdot 10^{-7}$ s$^{-1}$,
and $M_{Ni}$ is the total mass of the synthesized $^{56}$Ni. 
A fraction of the decay energy goes to generating positrons, which are
decelerated and annihilated by the ejecta (see e.g. \cite{capell}). The
energy input due to positrons is approximated as
\begin{equation}
E_{+} = 0.08 ~ M_{Ni} ~ S_{Co} 
\end{equation}
Thus, the total energy generation rate becomes the sum of the gamma ray and positron
energy productions.

The gamma-rays as well as positrons must interact with the ejecta
through multiple scattering in order to deposit their energy, which is assumed
to be fully thermalized and re-radiated as optical photons. The total amount of 
deposited energy at each time $t$ is approximated as
\begin{equation}
\epsilon = E_{\gamma} ~ (1 - \exp(- \tau_{\gamma})) ~+ ~
E_{+} ~ (1 - \exp(- \tau_{+}))
\end{equation}
where 
\begin{equation}
\tau_{\gamma, +} ~=~ \kappa_{\gamma, +} ~ \rho_0(t) ~ R(t) ~ x_0 ~ 
( 1 + { {1-x_0^{n-1}} \over {n-1}}) 
\end{equation}
is the total optical depth for gamma-rays and positrons (\cite{capell}).
The opacity $\kappa$ is assumed to be constant for both gamma-rays 
and positrons. This is a major simplification, but often used in SNe models 
(e.g. \cite{capell}; \cite{pinto}).

If the diffusion time is short, the emitted energy per second is approximately equal
to the rate of the deposited energy, i.e. $L_{opt} \approx \epsilon$. We have adopted
this approximation for calculating the bolometric luminosity 
(similarly to \cite{maeda}), instead of solving the time-dependent transport
problem for optical photons (see \cite{capell}). This condition  
is certainly not valid for pre-maximum epochs, thus, our simple light curve
model is applicable only on the declining part of the light curve. 

The light curve model outlined above contains the following free 
parameters: the total ejected mass $M$, the nickel mass $M_{Ni}$, 
the expansion velocity $v_{exp}$, the density exponent $n$, and 
the opacities $\kappa_{\gamma}$ and $\kappa_{+}$. These parameters
can be constrained by the observations. However, it is known that 
the light curve alone does not determine all parameters uniquely,
i.e. the same light curve could be described by several parameter
combinations (\cite{b5}; \cite{iwam}). Therefore, additional
(usually spectroscopic) data are needed to constrain the physical
parameters. Since our code is not capable of modeling the
spectrum, we have used the \ion{Si}{ii} velocities for this purpose.

In a grey atmosphere (such as our simple model with constant opacity) 
the location of the photosphere $x_{ph}$ is usually defined as
\begin{equation}
\int_{x_{ph}}^{1}{\kappa_{opt} ~ \rho(x,t) ~ R(t)~ dx} ~ = ~ {2 \over 3}
\end{equation}
Using Eq.2, the photospheric velocity can be expressed as
\begin{equation}
v_{ph} = v_{exp} ~ [1 - { {2(1-n)} \over 3} {1 \over {\kappa_{opt} ~
\rho_0(t) ~x_0^n ~ R(t)} } ]^{1/{1-n}}
\end{equation}

We have fitted Eq.16 and 19 to the bolometric light curve and the radial
velocity curve presented in Sect.2. Three kinds of models were defined
depending on the density configuration (characterized by the 
density parameters $x_0$ and $n$): a constant-density model with $n=0$ fixed
(Model A), a power-law model with $x_0 = 0.01$ fixed (Model B) and a
``core-shell'' model with $x_0 = 0.15$ fixed (Model C). In the latter two, 
the density exponent $n$ was treated as variable. In addition, 
the parameters $M$, $M_{Ni}$, $v_{exp}$, $\kappa_{\gamma}$, $\kappa_{+}$
and $\kappa_{opt}$ were varied to find the best agreement between the
observed and computed data. 

The fitted parameters are collected in Table 5. 
In general, low-mass models with $M = 1$ $M_{\odot}$ gave adequate
fit to the observed data. All models have shallow density dependence 
($n \approx 2$). The kinetic energy of the explosion 
$E_{51}$ is between 4 - 9 and the nickel mass is 
$M_{Ni} = $ 0.04 - 0.07 $M_{\odot}$. 
For Model C, we have also forced an $M = 2$ $M_{\odot}$
model fit (model C2 in Table 5). This model resulted in slightly worse,
but still acceptable fit (with less Ni mass, expansion velocity and opacity 
values). Similar tendency was found in the case of Models A and B: 
the increase of the ejected mass required the decrease of the Ni mass, velocity 
and opacities. It indicates that the parameter degeneracy, mentioned above,
is not completely solved by introducing the velocity data. 

Fig.11 and 12 show the comparison of Model C1 (solid line) and C2
(dashed line) with the observations. Model A and B resulted in very
similar light and velocity curves (note that the models are valid only
for the late-time light curve). 

Because only Model C could be 
fitted with the usual value of the gamma-opacity 
$\kappa_{\gamma} = 0.027$ cm$^2 g^{-1}$, we have adopted 
model C1 as the most probable. This model is somewhat similar
to the two-component model described by \cite{maeda}, because the
density structure and the optical depth is different for the core 
and the outer shell. \cite{maeda} discussed that a dense core 
is responsible for the late-time decline of the light curve.
Note that our simple constant density model (Model A) 
can also describe the late-time light curve, but this requires a radical 
increase of the gamma-ray and positron opacities, which is difficult
to explain physically. 

Indications for a shallow density profile of the ejecta was also 
given by spectral synthesis models. Using the parametrized
spectral synthesis code SYNOW, \cite{kinu} obtained 
$\tau \sim (v/v_{ph})^{-3}$ for the optical depth above the
photosphere. Assuming homologous expansion ($v \sim x$) and grey atmosphere
($\kappa$ constant), this corresponds to $\rho \sim x^{-4}$. 
The hydrodynamical calculations (\cite{mazz}) also resulted in a flat
density distribution above the photosphere. The result of our simple
model ($\rho \sim x^{-2}$) is consistent with these more sophisticated
calculations.
  
\begin{table}
\caption{Comparison of model parameters of SN~2002ap.}
\begin{tabular}{cccl}
\hline
\hline
$M_{ej}$ & $M_{Ni}$ & $E_{51}$ & Ref. \\
\hline
2.5 - 5 & 0.07 & 4 - 10 & \cite{mazz} \\
2.7 & 0.08 & 5.5 & \cite{maeda} \\
 - & 0.06 & - & \cite{pande} \\
1.5 & 0.09 & - & \cite{foley} \\
1.0 & 0.07 & 5.7 & Present paper \\
\hline
\end{tabular} 
\end{table}

\begin{figure}
\begin{center}
\leavevmode
\psfig{file=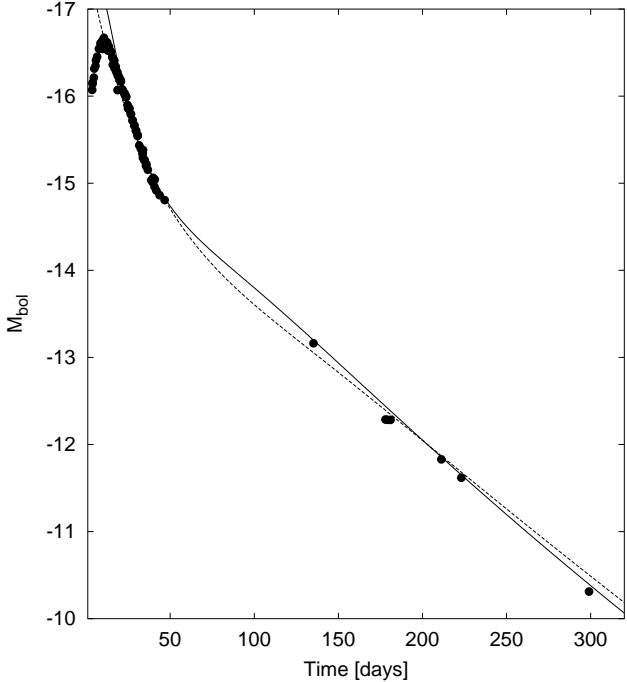,width=8.8cm}
\caption{Comparison of the computed bolometric light curves with the observations.
Continuous line: Model C1; dashed line: Model C2 (see Table 5 for the list of
parameters).}
\end{center}
\end{figure}

\begin{figure}
\begin{center}
\leavevmode
\psfig{file=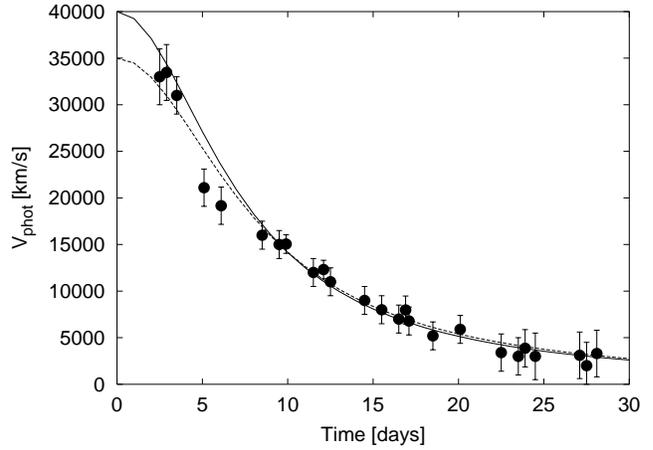,width=8.8cm}
\caption{Observed (dots) and computed photospheric radial velocities. 
The line types correspond to the same models as in Fig.11.}
\end{center}
\end{figure}

Table~6 summarizes the basic parameters $(M_{ej}, M_{Ni}, E_{51})$ of SN~2002ap 
collected from the literature. It is seen that our Model C1 has
lower ejected mass than the previous 
models, but the Ni-mass and the kinetic energy is similar. 
This is a bit surprising,
because we used a somewhat shorter distance modulus than the
authors in the previous papers, thus, our bolometric light curve
is slightly fainter. Because the peak luminosity is directly
connected with the Ni-mass, one should expect a lower Ni-mass
for a fainter peak luminosity. However, our light curve code is
probably too simplistic for such fine-tuning of the physical parameters
of SN 2002ap, thus, the parameters listed in Table 6 can be considered
only as order-of-magnitude estimates. As \cite{maeda} pointed out,
a two-component (or core-shell) ejecta model may, in fact, be
an indication of an asymmetric explosion. If this is indeed the case, the
parameters derived from such spherical models may deviate significantly
from the real values.   

We conclude that the simple model of the light- and velocity curves 
of SN~2002ap confirms that the progenitor was a low-mass object (probably 
a stripped stellar core), and the synthesized nickel mass was quite low, 
about 0.07 $M_{\odot}$, or even less, depending on the density structure
of the ejected envelope.

\section{Summary}

The results of this paper can be summarized as follows.

\begin{enumerate}

\item{We have presented new $BVRI$ photometry of SN 2002ap up to
$t = 300$ days past maximum, and photospheric phase optical spectroscopy.}

\item{The Expanding Photosphere Method was used to infer the distance of
SN 2002ap. This resulted in $d = 6.7$ Mpc (corresponding to $\mu_0 = 29.13$ mag)  
with $\pm 0.5$ Mpc random and $\pm 4$ Mpc (estimated) systematic uncertainty. 
This agrees with previous photometric distances of M74, 
if the reddening map of \cite{sfd} is used to estimate the galactic 
reddening. However, it is slightly less than the value (7.3 Mpc) 
used in previous papers of SN 2002ap.}

\item{The physical parameters of the SN ejecta were estimated with
a simple parametrized analytic model, assuming homologous expansion
and power-law density structure. From simultaneous fitting to the 
bolometric light curve and the radial velocity curve, we found that
the SN ejected a low mass envelope ($M_{ej} \approx 1.0 M_{\odot}$)
and synthesized about 0.04 - 0.07 $M_{\odot}$ of $^{56}$Ni. }

\end{enumerate}

\begin{acknowledgement}
This work was supported by Hungarian OTKA Grant No. T034615,
and the Bolyai J\'anos Research Scholarship to J.V. 
R.M.B. acknowledges financial support
from the Natural Science and Engineering Research Council through a
research grant to C. T. Bolton. 
The authors wish to express their thanks to the staff of Konkoly 
Observatory (director Prof. L. G. Bal\'azs) and David Dunlap Observatory 
(director Prof. S. M. Rucinski) for generously granting telescope time. 
Thanks are also due to Zs. Bebesi, Sz. M\'esz\'aros, 
P. Sz\'ekely, A. G\'asp\'ar, M. V\'aradi (University of Szeged)
and B. Sip{\H o}cz (ELTE University, Budapest) for assisting 
during the photometric observations. The authors are grateful to
an anonymous referee who provided many useful suggestions and 
comments that helped to improve this paper. 
The NASA Astrophysics Data System, the SIMBAD and NED databases
and the Canadian Astronomy Data Centre were used to access data
and references. The availability of these services is also 
gratefully acknowledged. 
\end{acknowledgement}


\begin{thebibliography}{}


\bibitem[Arnett, 1980]{arnett1}
   Arnett, W.D. 1980, ApJ 237, 541
   
\bibitem[Arnett, 1982]{arnett2}
   Arnett, W.D. 1982, ApJ 253, 785   

\bibitem[Barbon et al., 1990]{barbon}
   Barbon, R., Benetti, S., Rosino, L. et al. 1990, A\&A 237, 79

\bibitem[Berger et al., 2002]{b10}
   Berger, E., Kulkarni, S.R., Chevalier, R.A. 2002, ApJ 577, L5

\bibitem[Bessell, 2001]{bess}
   Bessell, M.S. 2001, PASP 113, 66

\bibitem[Borisov et al. 2002]{bori}
   Borisov, G., Dimitrov, D., Semkov, E. et al. 2002, IBVS No.
   5264.

\bibitem[Bottinelli et al. 1984]{bott}
   Bottinelli, L., Gouguenheim, L., Paturel, G., de Vaucoulers, G. 
   1984, A\&AS 56, 381

\bibitem[Burstein \& Heiles, 1982]{buhe}
   Burstein, D., Heiles, C. 1982, AJ 87, 1165 

\bibitem[Capellaro et al. 1997]{capell}
   Capellaro, E., Mazzali, P.A., Benetti, S. et al. 1997, A\&A 328, 203
   
\bibitem[Chornock et al. 2003]{b3}
   Chornock, R., Foley, R.J., Filippenko, A.V., Papenkova, M.,
   Weisz, D. 2003, IAU Circ. 8114
   
\bibitem[Contardo et al. 2000]{contardo}
   Contardo, G., Leibundgut, B., Vacca, W.D. 2000, A\&A 359, 876    

\bibitem[Cook et al. 2002]{cook}
   Cook, L.M., Katkova, E.V., Sokolov, N.A., Guseva, I.S. 2002,
   IBVS No. 5283

\bibitem[Eastman et al. 1996]{eastman}
    Eastman, R.G., Schmidt, B.P., Kirshner, R. 1996, ApJ 466, 911

\bibitem[Foley et al. 2003]{foley}
   Foley, R., Papenkova, M., Swift, B. et al. 2003, PASP 115, 1220

\bibitem[Galama et al. 1998]{b1}
   Galama T.J. et al. 1998, Nature 395, 670

\bibitem[Gal-Yam \& Shemmer, 2002]{b9}
   Gal-Yam, A., Shemmer, O. 2002, IAU Circ. 7811
  
\bibitem[Gal-Yam et al. 2002]{galy}
   Gal-Yam, A., Ofek, E.O., Shemmer, O. 2002, MNRAS 332, L73
   
\bibitem[Gal-Yam et al. 2004]{galy2}
   Gal-Yam, A., Moon, D.-S., Fox, D.B. et al. 2004, ApJ 609, 59  

\bibitem[Garnavich et al. 2003]{b2}
   Garnavich, P., Matheson, T., Olszewski, E.W., Harding, P., 
   Stanek, K.Z. 2003, IAU Circ. 8114
   
\bibitem[Hamuy et al. 2001]{hamuy}
    Hamuy, M., Pinto, P.A., Maza, J. et al. 2001, ApJ 558, 615 
    
\bibitem[Hamuy, 2003]{hamu2}
    Hamuy, M. 2003, ApJ 582, 905

\bibitem[H\"oflich, Wheeler \& Wang, 1999]{hoflich}
    H\"oflich, P., Wheeler, J.C., Wang, L. 1999, ApJ 521, 179

\bibitem[Iwamoto et al. 1998]{b5}
   Iwamoto, K. et al. 1998, Nature 395, 672
   
\bibitem[Iwamoto et al. 2002]{iwam}
   Iwamoto, K., Nomoto, K., Mazzali, P.A., Nakamura, T., Maeda, K. 2003,
   in: Supernovae and Gamma-Ray Bursts (ed. K.Weiler) Springer-Verlag p.243

\bibitem[Kawabata et al. 2002]{kawa}
   Kawabata, K.S., Jeffery, D.J., Iye, M. et al. 2002, ApJ 580, 39
   
\bibitem[Kinugasa et al. 2002]{b8}
   Kinugasa, K., Kawakita, H., Ayani, K. et al. 2002, IAU Circ.
   7811   

\bibitem[Kinugasa et al. 2002]{kinu}
   Kinugasa, K., Kawakita, H., Ayani, K. et al. 2002, ApJ 577, L97

\bibitem[Kirshner \& Kwan, 1974]{kirshner}
   Kirshner, R.P., Kwan, J. 1974, ApJ 193, 27

\bibitem[Leonard et al. 2002a]{leo}
   Leonard, D.C., Filippenko, A.V., Chornock R., Foley, R.
   2002, PASP 114, 1333
   
\bibitem[Leonard et al. 2002b]{leo2}
   Leonard, D.C., Filippenko, A.V., Chornock R., Li, W.
   2002, AJ 124, 2506

\bibitem[Leonard et al. 2003]{leo3}
   Leonard, D.C., Kanbur, S.M., Ngeow, C.C, Tanvir, N.R. 
   2003, ApJ 594, 247 
   
\bibitem[Lipkin et al., 2004]{lipkin}
   Lipkin, Y.M., Ofek, E.O., Gal-Yam, A. et al. 2004, ApJ 606, 381    
     
\bibitem[Maeda et al. 2003]{maeda}
   Maeda, K., Mazzali, P.A., Deng, J. et al. 2003, ApJ 593, 931
   
\bibitem[Mazzali, Iwamoto \& Nomoto, 2000]{maz97}
   Mazzali, P.A., Iwamoto, K., Nomoto, K. 2000, ApJ 545, 407   

\bibitem[Mazzali et al. 2002]{mazz}
   Mazzali, P.A., Deng, J., Maeda, K. et al. 2002, ApJ 572, L61

\bibitem[Meikle et al. 2002]{b7}
   Meikle, P., Lucy, L., Smartt, S. et al. 2002, IAU Circ. 7811

\bibitem[Munari \& Zwitter, 1997]{muzw}
   Munari, U, Zwitter, T. 1997, A\&A 318, 269

\bibitem[Nakano et al. 2002]{b6}
   Nakano, S., Kushida, R., Kushida, Y., Li, W. 2002, IAU Circ.
   7810

\bibitem[Pandey et al. 2002]{pande}
   Pandey, S.B., Anupama, G.C., Sagar, R. et al. 2003, MNRAS 
   340, 375

\bibitem[Patat et al. 2001]{patat}
   Patat, F. et al. 2001 ApJ 555, 900

\bibitem[Pinto \& Eastman, 2000]{pinto}
   Pinto, P.A., Eastman, R.G. 2000, ApJ 530, 744

\bibitem[Sandage \& Tammann, 1974]{st}
   Sandage, A.R., Tammann, G.A. 1974, ApJ 194, 559

\bibitem[Schlegel et al. 1998]{sfd}
   Schlegel, D., Finkbeiner, D., Davis, M. 1998, ApJ 500, 525

\bibitem[Schmidt et al. 1994]{schmidt}
   Schmidt, B.P., Kirshner, R.P., Eastman, R.G. et al. 1994, ApJ 432, 42

\bibitem[Sharina et al. 1996]{shari}
   Sharina, M.E., Karachentsev, I.D., Tikhonov, N.A. 1996, A\&AS 119, 499

\bibitem[Smartt et al. 2002]{smartt}
   Smartt, S.J., Vreeswijk, P.M., Ramirez-Ruiz, E. et al. 2002,
   ApJ 572, 147

\bibitem[Sohn \& Davidge, 1996]{sohn}
   Sohn, Y-J., Davidge, T.J. 1996, AJ 111, 2280   

\bibitem[Sohn \& Davidge, 1998]{sohn2}
   Sohn, Y-J., Davidge, T.J. 1998, AJ 115, 130

\bibitem[Soria et al. 2003]{soria}
    Soria, R., Pian, E., Mazzali, P.A. 2003, A\&A 413, 107

\bibitem[Stanek et al. 2003]{b4}
   Stanek, K.Z., Matheson, T., Garnavich, P.M.  et al. 2003, ApJ 591, 17
    
\bibitem[Sutaria et al. 2002]{b11}
   Sutaria, F.K., Chandra, P., Bhatnagar, S., Ray, A. 2002, 
   A\&A 397, 1011

\bibitem[Takada-Hidai et al. 2002]{taka}
   Takada-Hidai, M., Aoki, W., Zhao, G. 2002, PASJ 54, 899

\bibitem[Totani, 2003]{b12}
   Totani, T. 2003, ApJ 598, 1151


\bibitem[Yoshii et al. 2003]{yosh}
   Yoshii, Y., Tomita, H., Kobayashi, Y. et al. 2003, ApJ 592, 467

\bibitem[Vink\'o et al. 2001]{vinko1}
   Vink\'o, J., Cs\'ak, B., Csizmadia, Sz. et al. 2001, A\&A 372, 824


\bibitem[Vink\'o et al. 2003]{vinko2}
   Vink\'o, J., B\'{\i}r\'o, I.B., Cs\'ak, B. et al. 2003, A\&A 397, 115


\bibitem[Wang et al. 2002]{wang}
   Wang, L., Baade, D., H\"oflich, P., Wheeler, J.C. 2003, ApJ 592, 457


\end{thebibliography}
\end{document}